\documentstyle[prd,aps,epsfig,amssymb]{revtex}
\let\jnfont=\rm
\def\NPB#1,{{\jnfont Nucl.\ Phys.\ B }{\bf #1},}
\def\PLB#1,{{\jnfont Phys.\ Lett.\ B }{\bf #1},}
\def\EPJC#1,{{\jnfont Euro.\ Phys.\ J.\ C }{\bf #1},}
\def\PRD#1,{{\jnfont Phys.\ Rev.\ D }{\bf #1},}
\def\PRL#1,{{\jnfont Phys.\ Rev.\ Lett.\ }{\bf #1},}
\def\MPLA#1,{{\jnfont Mod.\ Phys.\ Lett.\ A }{\bf #1},}
\def\JPG#1,{{\jnfont J.\ Phys.\ G}{\bf #1},}
\def\CTP#1,{{\jnfont Commun.\ Theor.\ Phys.\ }{\bf #1},}

\begin{document}
\draft
\preprint{}
\title{SUSY-Induced Top Quark FCNC Processes at Linear Colliders}

\author{Junjie Cao$^{a,b}$,  Zhaohua Xiong $^{a,c}$  and Jin Min Yang $^b$}

\address{$^a$ CCAST (World  Laboratory), P.O.Box 8730, Beijing 100080, China}
\address{$^c$ Institute of Theoretical Physics, Academia Sinica, Beijing 100080, China}
\address{$^d$ Graduate School of Science, Hiroshima University, Hiroshima 937-6256, Japan}

\date{\today}

\maketitle

\begin{abstract}

In the Minimal Supersymmetric Model (MSSM) the hitherto unconstrained flavor mixing
between top-squark and charm-squark will induce the flavor-changing neutral-current
(FCNC) interaction between top quark and charm quark, which then lead 
to various processes at the next generation linear collider (NLC), i.e., the top-charm 
associated productions via $e^+ e^-$,  $e^- \gamma$ and $\gamma \gamma$ collisions 
as well as the top quark rare decays $t \to c V$ ($V=g$, $\gamma $ or $Z$).
All these processes involve the same part of the parameter space of the MSSM. 
Through a comparative analysis for all these processes at the NLC, we found that the best
channel to probe such SUSY-induced top quark FCNC is the top-charm associated production 
in $\gamma \gamma$ collision, which occurs at a much higher rate than $e^+ e^-$ or 
$e^- \gamma$ collision and may reach the detectable level for some part of the parameter 
space. Since the rates predicted by the Standard Model are far below the detectable
level,  the observation of such FCNC events would be a robust evidence of new
physics.
\end{abstract}

\pacs{14.65.Ha 12.60.Jv 11.30.Pb}

\section{\bf Introduction}
The study of flavor-changing neutral-current
(FCNC) interactions has been playing an important role in testing the
Standard Model (SM) and probing new physics beyond the SM.
As the most massive fermion in the SM, the top quark is naturally 
regarded to be more sensitive to new physics than other fermions.
In the SM, due to the GIM mechanism, the top quark FCNC interactions
are absent at tree level and extremely small at loop levels.
In some new physics models beyond the SM the top quark FCNC may be significantly 
enhanced. Searching for the top quark FCNC would be a good probe for new physics.

Intensive activities to explore the top quark FCNC couplings have been undertaken
in recent years.  On the experimental side, the CDF and D0 collaborations have reported
interesting bounds on the FCNC top quark decays \cite{cdfd0} from Run 1 experiment and will 
tighten the bounds from the on-going Run 2 experiment.  On the theoretical side, various FCNC 
top quark decays and top-charm associated productions at high energy colliders were extensively 
studied in the SM \cite{tcvh-sm,eetc-sm}, 
the Minimal Supersymmetric Model (MSSM) \cite{mssm-rev,tcv-mssm,tch-mssm,eetc-mssm}
and other new physics models \cite{other}. These studies showed that the SM predictions for 
such top quark FCNC processes are far below the detectable level and the MSSM can enhance 
them by several orders to make them potentially accessible at future collider experiments
\cite{tcv-at-collider}. 

Due to its rather clean environment, the next generation linear collider (NLC) will be
an ideal machine to probe new physics. In such a collider, in addition to $e^+ e^-$ collision,
we can also realize $\gamma \gamma$ collision and $e^- \gamma$ collision with the photon beams 
generated by the backward Compton scattering of incident electron- and laser-beams \cite{JLC}.
The SUSY induced top-charm FCNC will give rise to various processes at the NLC, i.e., the 
top-charm associated productions via $e^+ e^-$,  $e^- \gamma$ and $\gamma \gamma$ collisions 
as well as the top quark rare decays $t \to c V$ ($V=g$, $\gamma $ or $Z$).  It is noticable 
that some of these processes, like the top-charm associated productions in $\gamma \gamma$ 
or $e^- \gamma$ collision, have not been studied in the framework of the MSSM. The production 
in $\gamma \gamma$ collision 
may be more important than in $e^+ e^-$ collision studied in the literature \cite{eetc-mssm}. 
The reason is twofold. Firstly, the process 
$\gamma \gamma \to t \bar c$ is a good probe of new physics because it is essentially 
free of any SM irreducible background \cite{background}. Secondly, unlike the process 
$e^+ e^- \to t \bar c$, which is $s$-channel suppressed in high energy collisions, 
there are $t$- and $u$-channel contributions to $\gamma \gamma \to t \bar c$  and thus 
its cross section may be much larger at the NLC.  
It is also noticable that all these FCNC processes at the NLC involve the same part of the 
parameter space of the MSSM \footnote{Since the FCNC decay $t \to ch$ involves more parameters,
we do not include it in our analysis.}. Therefore, it is necessary to  perform a comparative analysis 
for all these processes to find out which process is best to probe the top quark FCNC. 
This is the aim of this article.

This paper is organized as follows. In Sec. II, we discuss the mixing between
top-squark and charm-squark and derive the FCNC interaction Lagrangian in SUSY-QCD.
In Sec. III, we calculate the $t \bar c$ productions in  $\gamma \gamma$ 
and $e^- \gamma$ colllisions induced by such FCNC SUSY-QCD interactions.  
Numerical results for these productions at the NLC are given in Sec. IV, 
with the comparison to $t \bar c$ production in  $e^+ e^-$ collision 
and various FCNC top quark decays. Finally, a brief discussion and conclusion 
can be found in Sec. V.

\section{Flavor mixing of top-squark and charm-squark and the induced FCNC}

Many popular SUSY models predict the flavor mixings of sfermions.
For the squark sector, despite of the possible strong constraints on the down
type squark flavor mixings from the low-energy experimental data \footnote{The down
type squark
flavor mixings in large $\tan\beta$ limit could enhance the FCNC $B$ decays 
by several orders \cite{xiong} and will face tests in the on-going $B$-factory 
experiments.}, the mixings between top-squark and charm-squark are subject to  
no strong low-energy constraints \cite{constr}.  

Such a mixing between top-squark and charm-squark is well motivated in low-energy 
supergravity models (SUGRA) \cite{duncan}. 
In these models squark mass matrices are diagonalized simultaneously as 
with the quark matrices at the Planck scale. 
But when the terms evolve down to low energy, this diagonality is violated 
by radiative corrections. Due to large top quark mass, at low energy 
it is found \cite{hikasa} that the mixing between top-squark and charm-squark 
may be significant.
Note that only $\tilde c_L$ mixes with top-squark while $\tilde c_R$ does not 
in the approximation of neglecting charm quark mass. As shown in \cite{hikasa},
the mixing between $\tilde c_L$ and  $\tilde t_L$ is most likely to
be large, which is proportional to a sum of some soft masses.      

Motivated by the above arguments, in our analysis we assume the existence
of the mixing between $\tilde c_L$ and  $\tilde t_L$ and parameterize the
mixing as $\delta_L M_{\tilde Q} M_{\tilde Q_1}$, where $M_{\tilde Q}$ 
($M_{\tilde Q_1}$) is the soft mass parameter for left-handed squark
of third (second) generation and  $\delta_L$ is a dimensionless     
parameter representing the mixing strength.  $\delta_L$ can be 
calculated in terms of other parameters in a given model like
mSUGRA  \cite{hikasa}. But in our calculation we retain it as
a free parameter in the range of $0\sim 1$ \cite{constr}.  

Considering the mixing between  $\tilde t_L$ and
$\tilde t_R$ and neglecting the mixing between  $\tilde c_L$ and 
$\tilde c_R$, we obtain the $ 3 \times 3 $ mass-square matrix 
in the basis $( \tilde{t}_L, \tilde{t}_R, \tilde{c}_L)$:
\begin{eqnarray}
{\cal L}_{mass} = \left ( \begin{array}{ccc}  \tilde t_L^* &  \tilde t_R^* &  \tilde c_L^*  \end{array}  \right)
 \left ( \begin{array}{ccc}
M_{\tilde{Q}}^2 +m_t^2+D_L           &  m_t X                       & \delta_L M_{\tilde Q} M_{\tilde Q_1} \\
m_t X                                &  M_{\tilde U}^2 + m_t^2+ D_R & 0  \\
\delta_L M_{\tilde Q} M_{\tilde Q_1} &  0                           & M_{\tilde{Q_1}}^2+m_c^2+D_L
           \end{array}  \right )  
 \left ( \begin{array}{c}\tilde t_L \\ \tilde t_R \\  \tilde c_L \end{array}  \right) 
\label{smass1}
\end{eqnarray}
where $D_L\equiv m_Z^2\cos(2\beta) (\frac{1}{2}-\frac{2}{3} s_W^2)$, 
$D_R=\frac{2}{3} m_Z^2\cos(2\beta) s_W^2$, $X=A_t +\mu \cot \beta$
and $M^2_{\tilde U}$ is the soft-breaking mass term for right-handed 
top-squark. $A_t$ is the coefficient of the trilinear term $H_2 \tilde Q \tilde U$
in soft-breaking terms, $\mu$ is the mixing mass parameter between
$H_1$ and $H_2$ in the superpotential and $\tan\beta=v_2/v_1$ is ratio of 
the vacuum expectation values of the two Higgs doublets. 

The mass-square matrix in Eq.(\ref{smass1}) can be diagonalized by a $3 \times 3 $ unitary matrix $V$,
which rotates the interaction eigenstates ($\tilde t_L$, $\tilde t_R$, $\tilde c _L$) into mass
eigenstates $\tilde{q}_{1,2,3}$. Such a rotation results in the FCNC 
in both the weak interaction sector and the strong SUSY-QCD sector.   
Of course, the latter will be dominant. So in our analysis we only consider
the FCNC in SUSY-QCD sector, given by
\begin{eqnarray} \label{fcnc}
{\cal L}_{FCNC}&=&-\sqrt{2} g_s T^a \left [ \bar{\tilde{g}}_a
(V^{\dagger}_{i1} P_L-V^{\dagger}_{i2} P_R) t
\tilde{q}_i^{\ast}+\bar{\tilde{g}}_a V^{\dagger}_{i3} P_L c
\tilde{q}_i^{\ast} \right ] ,
\end{eqnarray}
where $P_{L,R}=(1\mp \gamma_5)/2$.

\section{\bf Calculations}

The SUSY-QCD FCNC in Eq.(\ref{fcnc}) will induce the top-charm associated production
in  $\gamma \gamma$ collision, as shown in Fig.~1.  Throughout our calculations the
charge conjugate production channel, i.e., the production of $\bar t  c$, has also been
included.
Neglecting the charm quark mass, the amplitude of this  process is given by
\begin{eqnarray}
{\cal M}=2i\alpha\alpha_s Q_c^2 C_F ~\bar{u}_t\Gamma^{\mu\nu}P_Lv_c 
~\epsilon_\mu(\lambda_1)\epsilon_\nu(\lambda_2)
\end{eqnarray}
with $Q_c=2/3$, $C_F=4/3$ and 
\begin{eqnarray}
\Gamma^{\mu\nu}&=&
    c_1p_t^{\mu}p_t^{\nu}+c_2p_c^{\mu}p_c^{\nu}+c_3p_t^{\mu}p_c^{\nu}
   +c_4p_t^{\nu}p_c^{\mu} +c_5p_t^{\mu}\gamma^\nu+c_6p_c^{\mu}\gamma^\nu
   +c_7p_c^{\nu}\gamma^\mu+c_8p_t^{\nu}\gamma^\mu \nonumber\\
& &+c_9g^{\mu\nu}+c_{10}\gamma^{\mu}\gamma^{\nu}+c_{11}p_t^{\mu}p_t^{\nu} {\not{k}}_2
   +c_{12}p_c^{\mu}p_c^{\nu}{\not{k}}_2 +c_{13}p_t^{\mu}p_c^{\nu} {\not{k}}_2
   +c_{14}p_t^{\nu}p_c^{\mu}{\not{k}}_2 +c_{15}p_t^{\mu}\gamma^\nu {\not{k}}_2\nonumber\\
& &+c_{16}p_c^{\mu}\gamma^\nu {\not{k}}_2 +c_{17}p_c^{\nu}\gamma^\mu {\not{k}}_2
   +c_{18}p_t^{\nu}\gamma^\mu {\not{k}}_2 +c_{19}g^{\mu\nu} {\not{k}}_2
   +c_{20}i\varepsilon^{\mu\nu\alpha\beta}\gamma_\alpha k_{2\beta}.
\label{m-e}
\end{eqnarray}
Here $k_{1,2}$ denote the momentum of incoming photons and $p_{t,c}$ the momentum of
outgoing top and charm quarks. The coefficients $c_i$ can be obtained by a straightforward 
calculation of the diagrams shown in Fig.~\ref{fig:feyman1} and their expressions 
are presented in the Appendix.

%%%%%%%%%%%%%%%%%%%%%%%%%%%%%%%%%%%%%%%%%%%%%%%%%%%%%%%%%%%%%%%%%%%%%%%%%%%%%%%%%%%
\begin{figure}[tbh]
\begin{center}
\epsfig{file=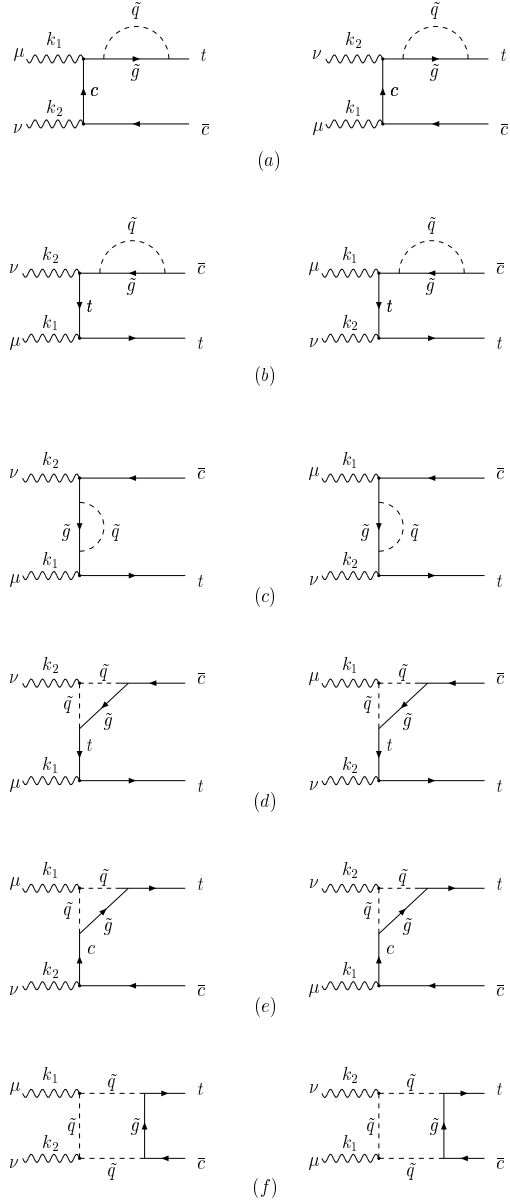,width=8cm} \caption{ Feynman diagrams contributing to
the process $\gamma\gamma \to t\bar c$ in SUSY-QCD.}
\label{fig:feyman1}
\end{center}
\end{figure}
%%%%%%%%%%%%%%%%%%%%%%%%%%%%%%%%%%%%%%%%%%%%%%%%%%%%%%%%%%%%%%%%%%%%%%%%%%%%%%%%%%%

We checked that our results satisfy the Ward identity, $k_1^\mu\Gamma_{\mu\nu}=0 $ and
$ k_2^\nu \Gamma_{\mu\nu}=0$. We also checked that all ultraviolet divergences canceled
in our results due to the unitary of $V$, which is essentially guaranteed by the 
renormalizability of the MSSM. 
Note that when $t$ or $u$ approaches 0, which is physically permitted,  $c_i$ tend to be 
very large. Indeed, this is the advantage of such a process over 
$e^+ e^- \to t \bar c $ which has only $s$-channel contribution and thus
is suppressed at high energy colliders.

The  $t \bar c $ production in  $e^- \gamma$ collision proceeds through the process
$e^- \gamma \to e^- \gamma^{\ast} \gamma  \to e^- t \bar c $, where the $\gamma$-beam 
is generated by the backward Compton scattering of incident electron- and laser-beam
and the $\gamma^{\ast}$ is radiated from $e^-$ beam. The subprocess 
$\gamma^{\ast} \gamma  \to t \bar c $ has the same Feynman diagrams shown in Fig.~\ref{fig:feyman1}.
In our calculation we use the  Weizs$\ddot{a}$cker-Williams approximation \cite{willam} 
which treats $\gamma^{\ast}$ from  $e^-$ beam as a real photon.
Thus the process can be approximated by the simpler fusion reaction
$\gamma \gamma \to  t \bar c $ and  its cross section is given by
\begin{eqnarray}
\hat \sigma_{e \gamma \to  e t \bar c}(s_{e \gamma})&=& \int_{(m_t+m_c)^2/s_{e\gamma}}^1  
{\rm d} x~P_{\gamma/e}(x,E_e)
~\hat{\sigma}_{\gamma \gamma \to  t \bar c}(s_{\gamma \gamma}=x s_{e \gamma}) ,
\end{eqnarray}
where $P_{\gamma/e} (x, E_e)$ is the probability of finding a photon with a fraction $x$ 
of energy $E_e$ in an ultrarelativistic electron and is given by \cite{willam}
\begin{eqnarray}
P_{\gamma/e}(x, E_e)=\frac{\alpha}{\pi} \left ( \frac{1+(1-x)^2}{x}
 \left ( \ln{\frac{E_e}{m_e}}-\frac{1}{2} \right )+
\frac{x}{2} \left ( \ln{(\frac{2}{x}-2)}+1 \right ) +
\frac{(2-x)^2}{2 x} \ln{\left ( \frac{2-2 x}{2-x} \right )} \right) .
\end{eqnarray}
Note that there are also intermediate $Z$-boson contribution for the process
$e \gamma \to e t \bar c$. However, such contributions are suppressed by
the probability function of finding a $Z$-boson in an ultrarelativistic electron \cite{effield}
and can be safely neglected.

For both $\gamma \gamma$ collider and $e \gamma$ collider, the photon beams are generated by the 
backward Compton scattering of incident electron- and laser-beams just before the interaction
point. The events number is obtained by convoluting the cross section with the photon beam 
luminosity distribution. For $\gamma \gamma$ collider the events number is obtained by
\begin{eqnarray}
N_{\gamma \gamma \to  t \bar c}&=&\int  {\rm d} \sqrt{s_{\gamma \gamma}}
\frac{{\rm d}{\cal L}_{\gamma \gamma}}
{{\rm d} \sqrt{s_{\gamma \gamma}}} \hat{\sigma}_{\gamma \gamma \to  t \bar c}
(s_{\gamma \gamma})\equiv {\cal L}_{ee} ~\sigma_{\gamma \gamma \to  t \bar c}(s_{ee}),
\label{definition}
\end{eqnarray}
where ${\rm d}{\cal L}_{\gamma \gamma}/{\rm d} \sqrt{s_{\gamma \gamma}}$ is the photon beam luminosity
distribution and $\sigma_{\gamma \gamma \to  t \bar c}(s_{ee}) $, with $s_{ee}$ being the 
energy-square of $e^+e^-$ collison, is defined as the effective cross
section of $ \gamma \gamma \to  t \bar c$.  In optimum case, it can be written as\cite{rrcl}
\begin{eqnarray}
\sigma_{\gamma \gamma \to  t \bar c}(s_{ee})&=&\int_{(m_t+m_c)/\sqrt{s_{ee}}}^{x_{max}} 2 z{\rm d} z
 ~\hat{\sigma}_{\gamma \gamma \to  t \bar c} (s_{\gamma \gamma}=z^2 s_{ee}) 
\int_{z^2/x_{max}}^{x_{max}} \frac{{\rm d} x}{x}~F_{\gamma/e}(x) ~F_{\gamma/e}(\frac{z^2}{x}),
\label{cross}
\end{eqnarray}
where $F_{\gamma/e}$ denotes the energy spectrum of the back-scattered photon for unpolarized initial electrons
and laser photon beams given by 
\begin{eqnarray}
F_{\gamma/e}(x)&=&\frac{1}{D(\xi)} \left ( 1-x+\frac{1}{1-x}-\frac{4 x}{\xi (1-x)}+
\frac{4 x^2}{\xi^2 (1-x)^2} \right ) .
\end{eqnarray}
The definitions of parameters $\xi$, $D(\xi)$ and $x_{max}$ can be found in \cite{rrcl}.
In our numerical calculation, we choose $\xi=4.8$, $D(\xi)=1.83$ and $x_{max}=0.83$.

For the $e^- \gamma $ collider the effective cross section of $e \gamma \to  e t \bar c$ 
is defined as
\begin{eqnarray}
\sigma_{e \gamma \to  e t \bar c}(s_{ee})&=& \frac{1}{{\cal L}_{ee}} \int {\rm d} \sqrt{s_{e\gamma}}
\frac{{\rm d}{\cal L}_{e \gamma}}
{{\rm d} \sqrt{s_{e \gamma}}} ~\hat \sigma_{e \gamma \to  e t \bar c} (s_{e \gamma}) \nonumber \\
&=& \int_{(m_t+m_c)/\sqrt{s_{ee}}}^{x_{max}} 2 z{\rm d} z
~\hat{\sigma}_{\gamma \gamma \to  t \bar c}
 (s_{\gamma \gamma}=z^2 s_{ee})
\int_{z^2/x_{max}}^{1} \frac{{\rm d} x}{x} 
 ~P_{\gamma/e}(x, E_e) F_{\gamma/e}(\frac{z^2}{x}).
\label{cross1}
\end{eqnarray}

The process  $e^+ e^- \to t \bar c$ and the top quark rare decays
$t \to  c V$ ($V=\gamma, Z, g$) were already calculated in the literature.
Here for comparison we recalculated all of them. The lengthy expressions
are not presented here.

\section{Numerical Results}
The relevant SUSY parameters  all the top-charm productions and top quark rare decays
are $\delta_L$, $X~(=A_t +\mu \cot \beta)$, $M_{\tilde g}$, $M_{\tilde Q}$, $M_{\tilde U}$,
$M_{\tilde Q1}$ and $\tan\beta$. To find out typical magnitudes of these processes, 
we scan over these SUSY parameters by requiring $ 5 \le \tan \beta \le 50 $,
$ m_{\tilde{q}} \ge 86.4$ GeV\cite{PDG00}, $ m_{\tilde{g}} >  190$ GeV \cite{PDG00},
$m_{\tilde{Q1}} \simeq 1$ TeV and restricting other soft mass parameters 
to be of sub-TeV scale. 

\begin{figure}[b]
\begin{center}
\epsfig{file=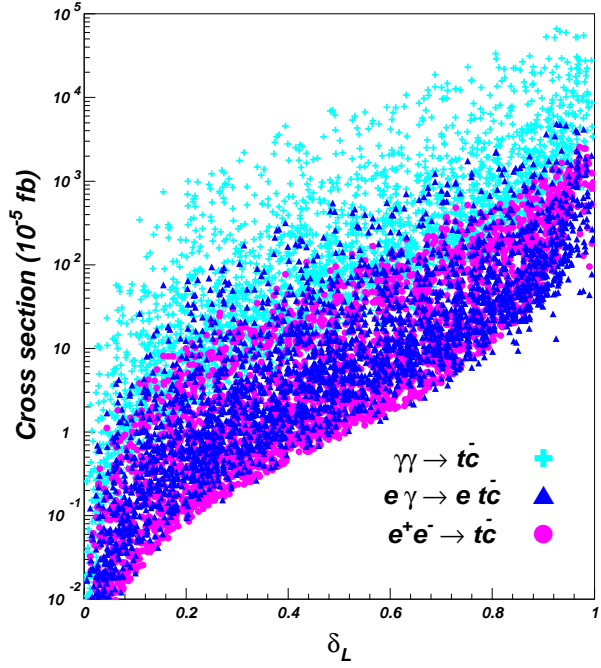,width=8.5cm}
\caption{The scattered  plot of  SUSY-QCD contribution to processes $\gamma \gamma \to  t \bar c$,
$ e \gamma \to  e t \bar c$ and $e^+ e^- \to t \bar c $ for $\sqrt{s_{ee}}=500~GeV$.}
\label{fig:scan1}
\end{center}
\end{figure}

\begin{figure}[b]
\begin{center}
\epsfig{file=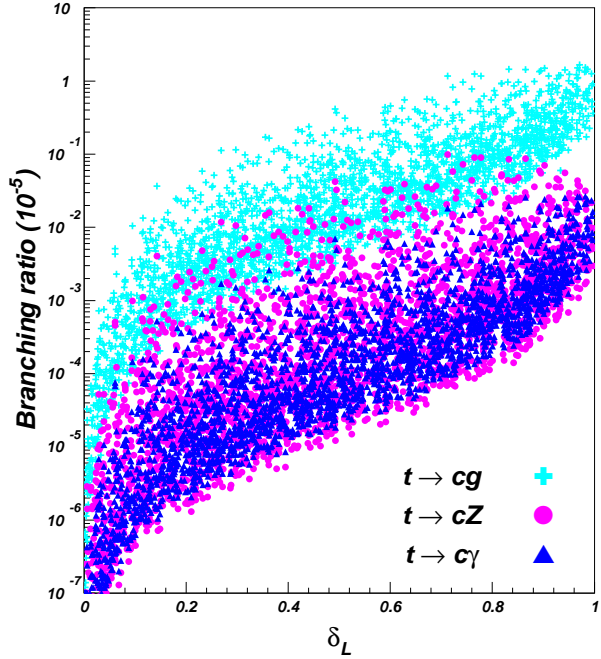,width=8.5cm}
\caption{Same as Fig.~\ref{fig:scan1}, but for FCNC top quark rare decays.}
\label{fig:scan2}
\end{center}
\end{figure}

Our scan results are plotted in Figs.~\ref{fig:scan1} and ~\ref{fig:scan2}.
For the productions one sees that the typical values are 
$\sigma (\gamma \gamma \to  t \bar c) \sim 10^{-2}$ fb and 
$\sigma (e \gamma \to  e t \bar c) \sim 10^{-3}$ fb and in optimum cases 
$\sigma (\gamma \gamma \to  t \bar c) \simeq 0.7$ fb and 
$\sigma (e \gamma \to  e t \bar c) \simeq 0.04$ fb. While for the process 
$e^+ e^- \to t \bar c$, the cross section can only reach $0.02$ fb, one order of 
magnitude lower than that of $\gamma \gamma \to t \bar c$ and comparable with that 
of $e \gamma \to e t \bar c $. The reason for this, as we pointed out before, is due 
to the $t$-channel enhancement for $\gamma \gamma \to t \bar c$ and $s$-channel
suppression for $e^+ e^- \to t \bar c$. 
For the top quark rare decays we see that the optimum values are $Br(t \to  c g)\sim 10^{-5}$,
$Br(t \to  c Z)\sim 10^{-7}$ and $Br(t \to  c \gamma)\sim 10^{-7}$, which agree with 
previous results \cite{tcv-mssm}.

\vspace*{-2cm}
%%%%%%%%%%%%%%%%%%%%%%%%%%%%%%%%%%%%%%%%%%%%%%%%%%%%%%%%%%%%%%%%%%%%%%%%%%%%%%%%%%%%%%%%%%%%%%%%
%%%%%%%%%%%%%%%%%%%%%%%%%%%%%%%%%%%%%%%%%%%%%%%%%%%%%%%%%%%%%%%%%%%%%%%%%%%%%%%%%%%%%%%%%%%%%%%%%
\vspace*{-1cm}
%%%%%%%%%%%%%%%%%%%%%%%%%%%%%%%%%%%%%%%%%%%%%%%%%%%%%%%%%%%%%%%%%%%%%%%%%%%%%%%%%%%%%%%%%%%%%%%%%
%%%%%%%%%%%%%%%%%%%%%%%%%%%%%%%%%%%%%%%%%%%%%%%%%%%%%%%%%%%%%%%%%%%%%%%%%%%%%%%%%%%%%%%%%%%%%

Comparing with predictions in other models,  such as R-parity violating model \cite{ma1} and 
the type III two Higgs doublet model\cite{ma2}, we find that the 
optimum value of $\sigma (\gamma \gamma \to t \bar c) $ in SUSY-QCD is larger.
Note that our results for $\sigma (e^+ e^- \to t \bar c)$ are quite different from those
in Ref.\cite{eetc-mssm} where the optimum value of $\sigma ( e^+ e^- \to t \bar c)$ can 
reach $0.1$ fb. The reason for the difference is that in Ref.~\cite{eetc-mssm} a non-unitary 
flavor changing matrix is used. During our scan, we find that, only in case of a light top-squark 
with mass of  $ 100 \sim 200 $ GeV and a large mass splitting of a few hundreds GeV 
in squark spectrum can relative large cross sections of these processes be obtained.
This is due to a GIM-like cancellation which occurs in the limit of degenerate 
squark masses as a result of the unitarity of the rotation matrix $V$.

The SM and MSSM predictions for the top-charm FCNC processes are summarized in Table I.  
Note that since the SM prediction for the process $\gamma \gamma \to  t \bar c$ has not been 
done in the literature, the SM value shown in the table is calculated by us. This is an arduous 
work. There are 66 diagrams contributing $\gamma \gamma \rightarrow t \bar c $ in 
$R_{\xi} $ gauge and often the contribution from a single box diagram can be decomposed into 
dozens of terms.  In order to avoid artificial error,  we use FormCalc\cite{formcalc} in
our calculations.
From Table I, we see that the MSSM can enhance the cross section of top-charm associated 
production processes by several orders.

\begin{figure}
\begin{center}
\epsfig{file=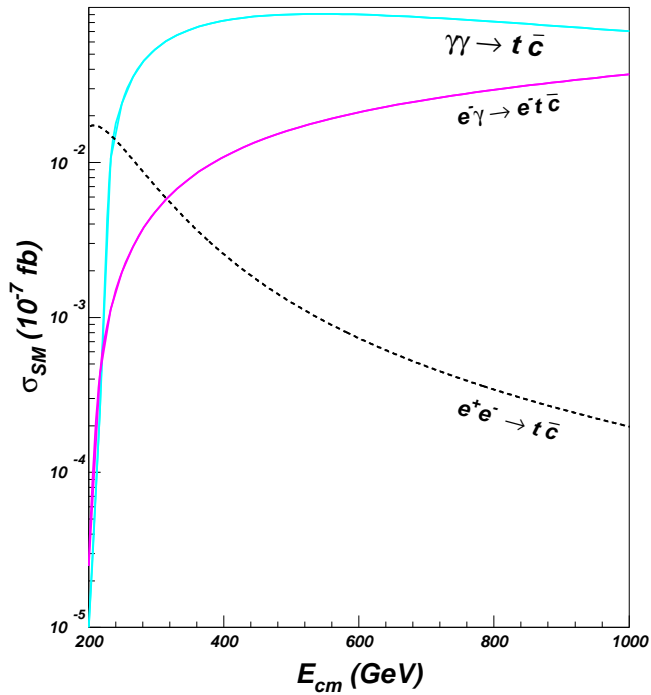,width=9cm}
\caption{Cross sections of top-charm associated production processes
versus collider energy $E_{cm}=\sqrt{s_{ee}}$ in the SM.}
\label{sm}
\end{center}
\end{figure}

%%%%%%%%%%%%%%%%%%%%%%%%%%%%%%%%%%%%%%%%%%%%%%%%%%%%%%%%%%%%%%%%%%%%%%%%%%%%%%%%%%%%%%%%%%%%%
\null
\noindent
{\small Table 1: Theoretical predictions for the top-quark FCNC
processes. SUSY-QCD predictions are the maximum values.
The collider energy is $500$ GeV for productions. }
\vspace*{0.1cm}
\begin{center}
\begin{tabular}{|l|l|l|}
\hline
 &~~~~SM&~~SUSY QCD~~  \\
\hline
~~$\sigma(\gamma\gamma \to t\bar c)$   &~~${\cal O}(10^{-8})$ fb  &~~${\cal O} (10^{-1})$ fb \\ \hline
~~$\sigma(e^- \gamma \to e^- t\bar c)$ &~~${\cal O}(10^{-9})$ fb  &~~${\cal O} (10^{-2})$ fb \\ \hline
~~$\sigma (e^+e^- \to t\bar c)$~~      &~~${\cal O}(10^{-10})$ fb &~~${\cal O}(10^{-2})$ fb \\ \hline
~~${\cal B}r (t \to cg)$               &~~${\cal O}(10^{-11})$    &~~${\cal O} (10^{-5})$ \\ \hline
~~${\cal B}r (t \to cZ)$               &~~${\cal O}(10^{-13})$    &~~${\cal O} (10^{-7})$ \\ \hline
~~${\cal B}r (t \to c \gamma)$         &~~${\cal O} (10^{-13})$   &~~${\cal O} (10^{-7})$ \\  \hline
\end{tabular}
\end{center}
%%%%%%%%%%%%%%%%%%%%%%%%%%%%%%%%%%%%%%%%%%%%%%%%%%%%%%%%%%%%%%%%%%%%%%%%%%%%%%%%%%%%%%%%%%%%%%%%%
\vspace*{.5cm}
The behaviours of the SM cross sections versus the collider energy are shown in Fig.~\ref{sm}
for three different production processes. We see that the cross section of $e^+e^- \to t\bar c$
drops quickly with the increase of collider energy due to the $s$-channel suppression.
 %%%%%%%%%%%%%%%%%%%%%%%%%%%%%%%%%%%%%%%%%%%%%%%%%%%%%%%%%%%%%%%%%%%%%%%%%%%%%%%
%%%%%%%%%%%%%%%%%%%%%%%%%%%%%%%%%%%%%%%%%%%%%%%%%%%%%%%%%%%%%%%%%%%%%%%%%%%%%%%

Let us focus on the most important process $\gamma \gamma  \to t \bar c $ 
and study some of its features. For example we fix $\tan \beta=30$ and $M_{\tilde{Q1}}=1000$ GeV,
and choose $ M_{\tilde Q}=600$ GeV, $M_{\tilde U}=400$ GeV for scenario I (relatively light squarks
for the third-family) and $ M_{\tilde Q}=800$ GeV, $M_{\tilde U}=600$ GeV for scenario II 
(relatively heavy squarks for the third-family). 

%%%%%%%%%%%%%%%%%%%%%%%%%%%%%%%%%%%%%%%%%%%%%%%%%%%%%%%%%%%%%%%%%%%%%%%%%%%%%%%
\begin{figure}
\begin{center}
\epsfig{file=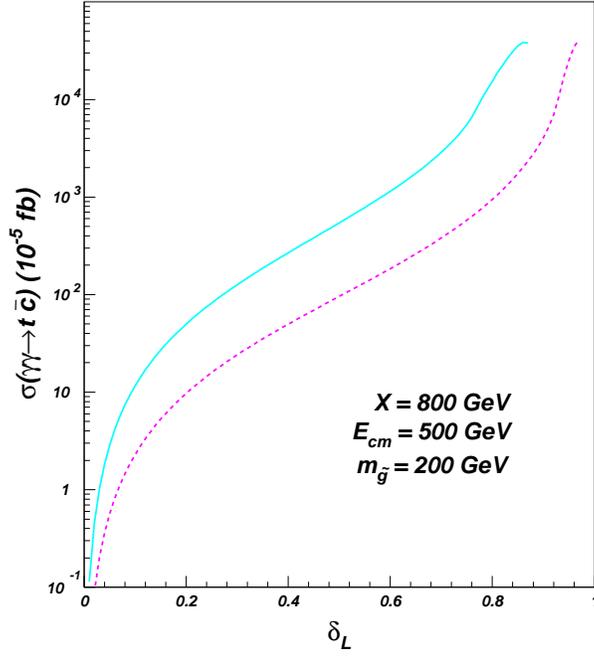,width=8.5cm} 
\caption{The SUSY-QCD correction to 
$\sigma_{\gamma \gamma  \to t \bar c }$ as a function of $\delta_L$. The solid 
curve corresponds to scenario I (relatively light squarks
for the third-family) and the dashed curve corresponds 
to scenario II (relatively heavy squarks for the third-family).}
\label{fig:sigma-del}
\end{center}
\end{figure}
%%%%%%%%%%%%%%%%%%%%%%%%%%%%%%%%%%%%%%%%%%%%%%%%%%%%%%%%%%%%%%%%%%%%%%%%%%%%%%%
%%%%%%%%%%%%%%%%%%%%%%%%%%%%%%%%%%%%%%%%%%%%%%%%%%%%%%%%%%%%%%%%%%%%%%%%%%%%%%%
\vspace*{-1cm}
\begin{figure}
\begin{center}
\epsfig{file=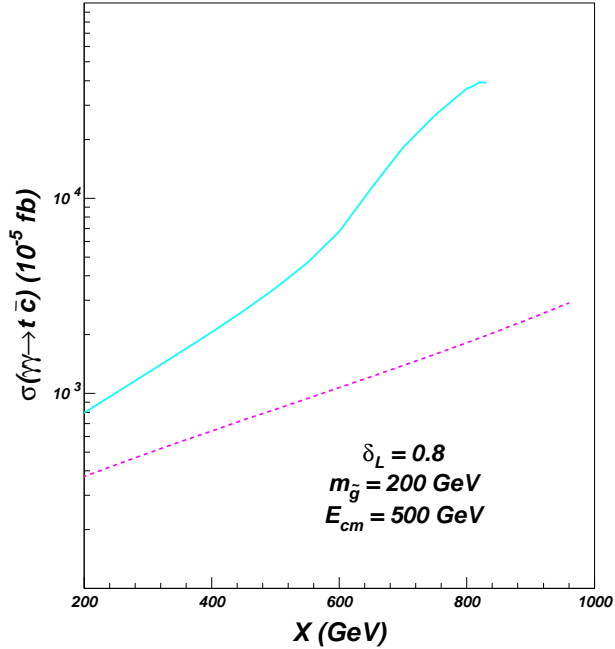,width=8.5cm}
\caption{Same as Fig~\ref{fig:sigma-del}, but for $\sigma(\gamma \gamma \to t \bar c)$ 
         as a function of $X~(=A_t +\mu \cot \beta)$.}
\label{fig:sigma-x}
\end{center}
\end{figure}
%%%%%%%%%%%%%%%%%%%%%%%%%%%%%%%%%%%%%%%%%%%%%%%%%%%%%%%%%%%%%%%%%%%%%%%%%%%%%%%%
 The cross section of $\gamma \gamma \to t \bar c$ as a function of  
$\delta_L$ is shown in Fig.~\ref{fig:sigma-del}. This figure shows 
that the cross section is enhanced dramatically with the increase of $\delta_L$. 
Clearly this is due to the fact that large $\delta_L$ will not only enhance the mixing
between top-squark and charm-squark but also enlarge the mass splitting between the squarks.
In Fig.~\ref{fig:sigma-x} we show the dependence of $\sigma(\gamma \gamma \to t \bar c)$ on $X$. 
We see that $\sigma(\gamma \gamma \to t \bar c)$ increases as $X$ gets large.
The reason is that large $X$ enhances the mass splitting between the top-squarks and 
thus lead to a weak cancellation between different Feynman diagrams.
%%%%%%%%%%%%%%%%%%%%%%%%%%%%%%%%%%%%%%%%%%%%%%%%%%%%%%%%%%%%%%%%%%%%%
\begin{figure}
\begin{center}
\epsfig{file=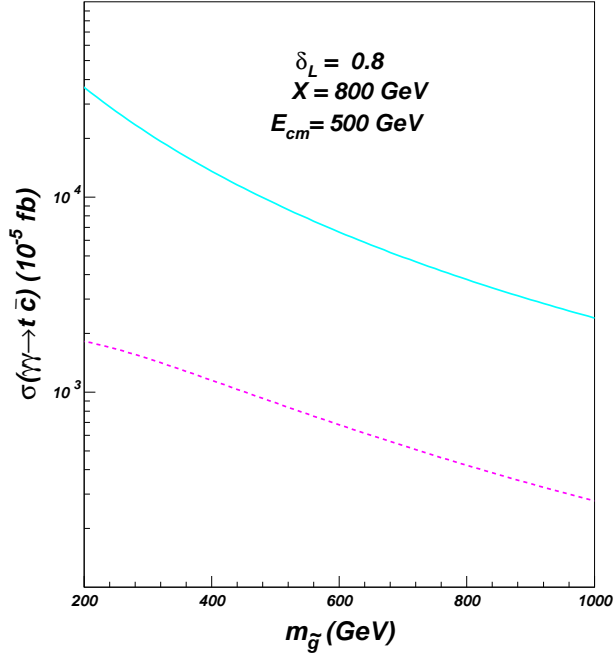,width=8.5cm} 
\caption{Same as Fig~\ref{fig:sigma-del}, but for the dependence on luino mass. } 
\label{fig:sigma-mg}
\end{center}
\end{figure}
%%%%%%%%%%%%%%%%%%%%%%%%%%%%%%%%%%%%%%%%%%%%%%%%%%%%%%%%%%%%%%%%%%%%%%
%%%%%%%%%%%%%%%%%%%%%%%%%%%%%%%%%%%%%%%%%%%%%%%%%%%%%%%%%%%%%%%%%%%%%%
\vspace*{-1cm}
\begin{figure}
\begin{center}
\epsfig{file=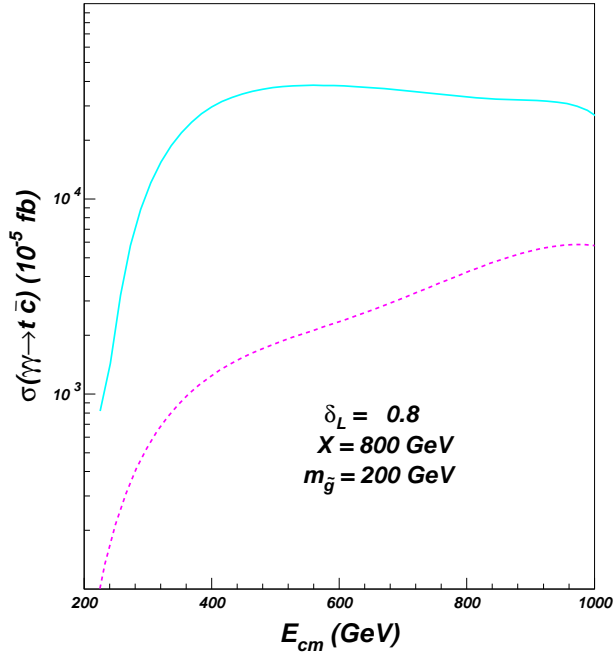,width=8.5cm}
\caption{Same as Fig~\ref{fig:sigma-del}, but for the dependence on $\sqrt{s} $. }
\label{fig:sigma-s}
\end{center}
\end{figure}
%%%%%%%%%%%%%%%%%%%%%%%%%%%%%%%%%%%%%%%%%%%%%%%%%%%%%%%%%%%%%%%%%%%%%%
The dependence of $\sigma(\gamma \gamma \to t \bar c)$ on $m_{\tilde{g}} $ and $\sqrt{s}$ 
is shown in Fig.~\ref{fig:sigma-mg} and Fig.~\ref{fig:sigma-s}, respectively.
Fig.~\ref{fig:sigma-mg} shows that the cross section drops when gluino becomes heavier, 
showing the decoupling effect of the MSSM. 

From Figs.~\ref{fig:sigma-del}-\ref{fig:sigma-s} one finds that the cross section 
in scenario I is always larger than that in scenario II.
The reason is that scenario I has the relative light squarks for the third-family which
lead to a larger mass splitting in squark spectrum.
%%%%%%%%%%%%%%%%%%%%%%%%%%%%%%%

\section{Summary and Conclusion}

We have performed a comparative analysis for the SUSY-induced 
top-charm FCNC processes at the NLC. These  processes include the top-charm 
associated productions via $e^+ e^-$,  $e^- \gamma$ and $\gamma \gamma$ collisions 
as well as the top quark rare decays $t \to c V$ ($V=g$, $\gamma $ or $Z$), all
of which involve the same SUSY parameters. 

In the production channels the $t \bar c$ production in $\gamma \gamma$ collision 
was found to occur at a much higher rate than $e^+ e^-$ or $e^- \gamma$ collision.
In some part of parameter space, the production rate of $\gamma \gamma \to t \bar c$ 
can reach $0.7$ fb. This means we may have $70$ events each year for the designed luminosity of
$100$ fb$^{-1}$/year at the NLC. Since the SM value of the production rate is 
completely negligible, the observation of such $t\bar c$ events would be a robust
indirect evidence of SUSY. 

Note that in practical experimental searches of such
$t\bar c$ events, a careful study of backgrounds is needed. To efficiently suppress
the  backgrounds, the signal events may be hurt by $50\%$ or more \cite{background,effect2}.
Using polarization at a linear collider can help to suppress the background, as analyzed
in the second reference of \cite{tcv-at-collider}.

In the rare decay channels the  $t \to c g$ was found to have the largest branching
ratio which can reach $10^{-5}$, in agreement with previous studies. 
Although $10^5$ $t\bar t$ events could be produced at the NLC each year,
studies \cite{bran} showed that the sensitivity to such rare decays  
can only reach $5\times 10^{-4}$. So the MSSM prediction for the branching
ratio of $10^{-5}$ is too low to be accessible at the NLC unless the
designed luminosity can be further upgraded.
             
Therefore, we conclude that, to probe the SUSY-induced  FCNC top quark interactions
at the NLC, $t \bar c$ production in $\gamma \gamma$ collision 
is the best channel.

Note that in our analysis we assumed
flavor mixing occurs between $\tilde c_L$ and  $\tilde t_L$, which is favored in SUGRA
models. Other kinds of mixings like the mixing between $\tilde c_R$ and  
$\tilde t_L$ \cite{cpyuan,atwood} may also be of phenomenological interest. 
However, we believe our conclusion will be quite model-independent since our results 
reflect the basic features of the processes. 

\section*{Acknowledgment}

We thank Kaoru Hagiwara and Tao Han for very helpful discussions. This work is supported by Young
Outstanding Foundation of Academia Sinica. 

\section*{Appendix}

Before presenting the explicit form of $\Gamma^{\mu \nu}$, we define the following abbreviations
\begin{eqnarray}
\hat{s}&=&(k_1+k_2)^2=(p_t+p_c)^2, \\
\hat{t}&=&(p_t-k_2)^2=(k_1-p_c)^2, \\
\hat{u}&=&(p_t-k_1)^2=(k_2-p_c)^2, \\
\hat{V}_{13} B^a_i&=&\sum_{\beta=1}^3 V_{1 \beta} V^{\dagger}_{\beta 3} B_i(-p_t, m_{\tilde{U}_{\beta}}, m_{\tilde{g}}), \\
\hat{V}_{13} B^b_i&=&\sum_{\beta=1}^3 V_{1 \beta} V^{\dagger}_{\beta 3} B_i(p_c, m_{\tilde{U}_{\beta}}, m_{\tilde{g}}), \\
\hat{V}_{13} B^c_i&=&\sum_{\beta=1}^3 V_{1 \beta} V^{\dagger}_{\beta 3} B_i(-p_t+k_2, m_{\tilde{U}_{\beta}}, m_{\tilde{g}}),\\
\hat{V}_{13} C^e_{i j}&=&\sum_{\beta=1}^3 V_{1 \beta} V^{\dagger}_{\beta 3} C_{i j}(k_1, p_c-k_1, m_{\tilde{U}_{\beta}}, m_{\tilde{U}_{\beta}}, m_{\tilde{g}}) ,\\
\hat{V}_{13} C^d_{i j}&=&\sum_{\beta=1}^3 V_{1 \beta} V^{\dagger}_{\beta 3} C_{i j}(k_2, -p_t, m_{\tilde{U}_{\beta}},
 m_{\tilde{U}_{\beta}}, m_{\tilde{g}}) ,\\
\hat{V}_{13} D^1_{i j}&=&\sum_{\beta=1}^3 V_{1 \beta} V^{\dagger}_{\beta 3} D_{i j}(k_2, -p_t, -p_c, m_{\tilde{U}_{\beta}},m_{\tilde{U}_{\beta}}, m_{\tilde{g}}, m_{\tilde{U}_{\beta}}), 
\end{eqnarray}
where $V$ is squark mixing matrix and B, C and D are loop functions defined in \cite{hooft}.

Then $\Gamma^{\mu \nu} $ is given by
\begin{eqnarray}
\Gamma^{\mu \nu}=\Gamma^{\mu \nu}_t+\Gamma^{\mu \nu}_u+\Gamma^{\mu \nu}_{Box1}+\Gamma^{\mu \nu}_{Box2}  \label{amp}
\end{eqnarray}
with 
$\Gamma^{\mu \nu}_u=\Gamma^{\mu \nu}_t |_{(k_1 \leftrightarrow k_2, \mu \leftrightarrow \nu,  t \leftrightarrow u})$
and 
$\Gamma^{\mu \nu}_{Box2}=\Gamma^{\mu \nu}_{Box1} |_{(k_1 \leftrightarrow k_2, \mu \leftrightarrow \nu,  t \leftrightarrow u})$.   $ \Gamma^{\mu \nu}_t $ and $\Gamma^{\mu \nu}_{Box1}$ in Eq.(\ref{amp}) can be expressed in the form of Eq.(\ref{m-e}) and their corresponding coefficients $c_i$ are given as
\begin{eqnarray}
c_{t,1}&=& \frac{4}{\hat{t}-m_c^2} (m_t \hat{V}_{13} (C^d_{22}-C^d_{23})-m_{\tilde{g}} \hat{V}_{23} C^d_{12})   ,  \\
c_{t,4}&=& \frac{-4}{\hat{t}-m_t^2} (m_t \hat{V}_{13} (C^e_{23}-C^e_{22})+m_{\tilde{g}} \hat{V}_{23} C^e_{12})+c_{t,1} ,    \\
c_{t,5}&=& \frac{-1}{m_t^2 (\hat{t}-m_c^2)} (m_t^2  \hat{V}_{13} (B^a_0+B^a_1) -m_{\tilde{g}} m_t  \hat{V}_{23} B^a_0) ,\nonumber \\  
& & -\frac{1}{m_t^2 (\hat{t}-m_t^2)} m_{\tilde{g}} m_t \hat{V}_{23} B^b_0-  \frac{1}{(\hat{t}-m_t^2) (\hat{t}-m_c^2)} ,  \nonumber  \\
&& \times (\hat{t} \hat{V}_{13}
 (B^c_0+B^c_1)-m_{\tilde{g}} m_t  \hat{V}_{23} B^c_0)  \nonumber \\
&&+\frac{2}{\hat{t}-m_c^2} \hat{V}_{13} C^d_{24}+
\frac{2}{\hat{t}-m_t^2} \hat{V}_{13} C^e_{24} ,  \\
c_{t,6}&=&-2 \hat{V}_{13} (C^e_{23}-C^e_{22})+c_{t,5} , \\
c_{t,8}&=&  \frac{2}{m_t^2 (\hat{t}-m_c^2)} (m_t^2  \hat{V}_{13} (B^a_0+B^a_1)-m_{\tilde{g}} m_t  \hat{V}_{23} B^a_0)  \nonumber \\
 && +\frac{2}{m_t^2 (\hat{t}-m_t^2)} m_{\tilde{g}} m_t \hat{V}_{23} B^b_0+ \frac{2}{(\hat{t}-m_t^2) (\hat{t}-m_c^2)} \nonumber \\
&& \times  (\hat{t} \hat{V}_{13}
 (B^c_0+B^c_1)-m_{\tilde{g}} m_t \hat{V}_{23} B^c_0)  \nonumber \\
& & -\frac{2}{\hat{t}-m_c^2} (\hat{t} \hat{V}_{13} (C^d_{12}+C^d_{23})+m_t^2 \hat{V}_{13} (C^d_{22}-C^d_{23})  \nonumber \\
& & +2 \hat{V}_{13} C^d_{24} - m_{\tilde{g}} m_t \hat{V}_{23} C^d_{12})
    -\frac{4}{\hat{t}-m_t^2} \hat{V}_{13} C^e_{24}  , \\
 c_{t,10}&=&\frac{-m_t}{m_t^2 (\hat{t}-m_c^2)} (m_t^2  \hat{V}_{13} (B^a_0+B^a_1)-m_{\tilde{g}} m_t  \hat{V}_{23} B^a_0) \nonumber \\
&&-\frac{m_t}{(\hat{t}-m_t^2) (\hat{t}-m_c^2)} (\hat{t} \hat{V}_{13} (B^c_0+B^c_1)
  -m_{\tilde{g}} m_t  \hat{V}_{23} B^c_0)  \nonumber \\
& & + \frac{2 m_t}{\hat{t}-m_c^2} \hat{V}_{13} C^d_{24} 
    +\frac{\hat{t}}{(\hat{t}-m_t^2) (\hat{t}-m_c^2)} (m_t \hat{V}_{13} (B^c_0+B^c_1)-m_{\tilde{g}} \hat{V}_{23} B^c_0)  , \\
 c_{t,16}&=& \frac{2}{\hat{t}-m_t^2} (m_t \hat{V}_{13} (C^e_{23}-C^e_{22})+m_{\tilde{g}} \hat{V}_{23} C^e_{12})   , \\
c_{t,18}&=& \frac{-2}{\hat{t}-m_c^2} (m_t \hat{V}_{13} (C^d_{22}-C^d_{23})-m_{\tilde{g}} \hat{V}_{23} C^d_{12})   ,  \\
c_{t, 19}&=&-c_{t, 20}=-c_{t,5} ,   \\
c_{Box1, 1}&=&4 (m_t \hat{V}_{13}-m_{\tilde{g}} \hat{V}_{23}) (D_{22}-D_{24}) 
              +4 m_t \hat{V}_{13} (D_{32}-D_{36})  , \\
c_{Box1, 2}&=&4 (m_t \hat{V}_{13}-m_{\tilde{g}} \hat{V}_{23}) (D_{23}-D_{25}) 
              -4 m_t \hat{V}_{13} (D_{310}-D_{39})  , \\
c_{Box1, 3}&=&-4 (m_t \hat{V}_{13}-m_{\tilde{g}} \hat{V}_{23}) (D_{25}-D_{26}) 
               -4 m_t \hat{V}_{13} (D_{310}-D_{37})  , \\
c_{Box1, 4}&=&-4 (m_t \hat{V}_{13}-m_{\tilde{g}} \hat{V}_{23}) (D_{24}-D_{26})
              -4 m_t \hat{V}_{13} (D_{36}-D_{37})  , \\
c_{Box1, 5}&=&-4 \hat{V}_{13} (D_{311}-D_{312})  , \\
c_{Box1, 6}&=&-4 \hat{V}_{13} (D_{311}-D_{313})  , \\
c_{Box1, 7}&=&4 \hat{V}_{13} D_{313}   , \\
c_{Box1, 8}&=&4 \hat{V}_{13} D_{312}  , \\
c_{Box1, 9}&=&-4 (m_t \hat{V}_{13}-m_{\tilde{g}} \hat{V}_{23}) D_{27} 
              +4 m_t \hat{V}_{13} D_{312}  , \\
c_{Box1, 11}&=&-4 \hat{V}_{13} (D_{22}-D_{24}-D_{34}+D_{36})  , \\
c_{Box1, 12}&=&-4 \hat{V}_{13} (D_{23}-D_{25}-D_{35}+D_{38})  , \\
c_{Box1, 13}&=&4 \hat{V}_{13} (D_{25}-D_{26}-D_{310}+D_{35})  , \\
c_{Box1, 14}&=&4 \hat{V}_{13} (D_{24}-D_{26}-D_{310}+D_{34})  , \\
c_{Box1, 19}&=&4 \hat{V}_{13} (D_{27}-D_{311}) . 
\end{eqnarray}

%\end{thebibliography}
\end{document}